\documentclass[aps, prl,
twocolumn,showpacs]{revtex4}
\usepackage{t1enc}
\usepackage[latin1]{inputenc}
\usepackage[english]{babel}
\usepackage{graphics}
\usepackage{graphicx}
\usepackage{dcolumn}
\usepackage{bm}
\usepackage{color}
\begin{document}
\title{Charge order in La$_{1.8-x}$Eu$_{0.2}$Sr$_x$CuO$_4$ studied by resonant soft X-ray diffraction}
\author{J.\ Fink,$^{1,2}$  E.\ Schierle,$^3$ E.\ Weschke,$^3$  J.\ Geck,$^4$ D.\ Hawthorn,$^4$ H.\ Wadati,$^4$ H.-H.\ Hu,$^5$  H. A.\ D\"urr,$^1$ N.\ Wizent,$^2$ B.\ B\"uchner,$^2$ G.A.\ Sawatzky,$^4$ }
\affiliation{
$^1$ BESSY, Albert-Einstein-Strasse 15,12489 Berlin, Germany\\
$^2$ Leibniz-Institute for Solid State and Materials
Research Dresden, P.O.Box 270116, D-01171
Dresden, Germany\\
$^3$Hahn-Meitner-Institut Berlin c/o BESSY, Albert-Einstein-Str. 15, D-12489 Berlin, 
Germany\\
$^4$Department of Physics and Astronomy,University of British Coulombia, 6224 Agricultural Road, Vancouver, BC, 
V6T 1Z1 Canada\\
$^5$II. Physikalisches Institut, Universit\"at zu K\"oln, Zülpicher Strasse 77, D-50937 K\"oln, Germany}

\date{\today}

\begin{abstract}
Resonant soft X-ray scattering with photon energies near the O $K$ and the Cu $L_3$ edges was used to 
study charge ordering in the system La$_{1.8-x}$Eu$_{0.2}$Sr$_x$CuO$_4$ as a 
function of temperature for x = 0.125 and  0.15. From the superstructure diffraction 
intensities a charge ordering with a doping dependent wave vector is derived which is in this system 
well below the transition temperature of the low-temperature tetragonal phase but well above the onset 
of spin ordering. This indicates that charge ordering is the 
primary driving force for the formation of stripe-like phases in two-dimensional doped cuprates.  Analysis of 
the lineshape of the scattered intensity as a function of photon energy yields evidence for a high
hole concentration in the stripes.
\end{abstract}
\maketitle
In the doped cuprates there exists a complex 
interplay between lattice, charge and spin degrees of freedom yielding several phases in a narrow 
concentration range: the antiferromagnetic insulating phase, the high-$T_c$ superconducting 
phase, the pseudo-gap region and charge- and spin-ordered phases. The latter are usually divided into 
checkerboard phases and stripe-like phases. The stripe-like phase in which antiferromagnetic antiphase magnetic 
domains are separated by periodically spaced domain walls to which the holes segregate was first predicted 
from Hartree-Fock analysis of the single-band Hubbard model \cite{Zaanen1989}. Fluctuating stripe-like order of 
spins in La$_{2-x}$Sr$_x$CuO$_4$ (LSCO) near x = 1/8 has been inferred 
from neutron scattering showing incommensurate magnetic 
correlations \cite{Birgeneau1989}. In La$_{1.875}$Ba$_{0.125}$CuO$_4$ (LBCO) superconductivity is 
strongly suppressed and a low-temperature tetragonal (LTT) 
phase appears which stabilizes a static stripe-like order due to the 
corrugated pattern of the in-plane lattice potential \cite{Fujita2002}.  A static stripe-like order  of the Cu 
spins has  also been detected in La$_{1.6-x}$Nd$_{0.4}$Sr$_x$CuO$_4$ (LNSCO) by neutron 
scattering. In these compounds the corrugation of the CuO$_2$ plane is more pronounced due to the smaller ionic 
radius of Nd compared to La \cite{Tranquada1995}. Replacing Nd by even smaller 
Eu ions, a phase diagram of the system La$_{1.8-x}$Eu$_{0.2}$Sr$_x$CuO$_4$ (LESCO) has been 
proposed based on $\mu$SR experiments \cite{Klauss2000}. There the magnetic stripe order almost completely replaces 
the superconducting range between x = 0.08 and 0.2. Only above x $\approx$ 0.2 
superconductivity could be detected in a narrow concentration range.
 
Using neutron scattering \cite{Tranquada1995} and non-resonant X-ray scattering \cite{Von1998}, the 
ordering of the charges can be only monitored indirectly by the associated lattice distortion.  
The reason for this is that neutron
and X-ray diffraction are mainly sensitive to the nuclear scattering and the core electron scattering, respectively. 
More direct information on the
charge modulation in cuprates can be obtained by resonant soft 
X-ray scattering (RSXS) using photon energies at the O $K$ and the Cu $L$ 
edge \cite{Abbamonte2002, Abbamonte2005}. In particular at the threshold of the O $K$ level the form factor 
for a charge carrier is enhanced by a factor of 82 \cite{Abbamonte2005}.
Near the Cu $L$ edge there is a strong enhancement of the form factor as well, but mainly lattice distortions
are probed which may be caused by charge modulations \cite{Abbamonte2006}. 
In the high-T$_c$ superconductor LSCO (x = 0.15) no static stripe-like modulation of the charge carriers
could be detected by RSXS \cite{Abbamonte2002} while in the compound LBCO (x = 1/8) a strong modulation
of the charge carriers has been observed in the range below the LTT phase transition temperature
T$_{LTT}$ = 55 K. We emphasize a striking feature of all investigations of stripe-like charge and spin order  
in this system: the ordering seems to be of cooperative nature, i.e. charge and spin order
show similar temperature dependence and the appearance of the charge order at T$_{CO}$ 
and a spin ordered state at T$_{SO}$ is found very close to T$_{LTT}$ \cite{Fujita2002}.
In LNSCO with x = 0.125 and T$_{LTT}$ = 70 K, on the other hand, lattice 
distortions due to charge ordering have been detected at T$_{CO}$ = 62 K 
and spin order at T$_{SO}$ = 54 K \cite{Tranquada1996}. The three transition temperatures 
in this system are slightly separated
and there is an indication that charge order may be more important for the formation of a stripe-like
order than spin order. In the compounds LESCO, which is the focus of this study, T$_{LTT}$ = 125 K and T$_{SO}$ = 45 
K, derived from neutron scattering \cite{Hucker2007}, are distinctly
different. The possible appearance of a separate charge and spin ordering should therefore be readily observable in this system.

In this contribution we have determined charge ordering directly by means of RSXS 
in a second system besides LBCO, namely in LESCO. 
From a lineshape analysis 
of the superstructure diffraction intensity as a function of photon energy across the O $K$ resonance, we infer 
the existence of a high doping concentration per Cu site in the hole stripes.
Furthermore, it is shown that the incommensurate charge ordering wave vector
exhibits a doping dependence  corresponding to that observed for spin order. 
In contrast to the previous RSXS study of LBCO, we find a modulation of the charge carrier density
 in a large temperature range without spin order. These findings clearly
show that the primary driving force for stripe formation is the charge channel.

The RSXS experiments were performed at the BESSY undulator beam line UE 46-PGM 
operated by the Hahn-Meitner Institut Berlin, working with vertically polarized photons in a horizontal
scattering geometry. A two-circle UHV 
diffractometer was equipped with a continuous flow He cryostat. 
The detector was a silicon diode which had an angular acceptance of 0.8° in the scattering plane 
and 4° perpendicular to it. 
LESCO single crystals were grown using the traveling solvent floating zone method. 
X-ray diffraction, $^{63}$Cu NMR spectroscopy and magnetic susceptibility measurements yielded for 
both concentrations a first order phase transition at T$_{LTT}$ = 125 K \cite{Simovic2003}.
The single crystals were cleaved parallel to the 
(001) surface in air shortly before the transfer to the scattering chamber. 
At a photon energy of 1100 eV the (002) peak could be reached which was used together with the superstructure peaks to orient 
the sample. 

\begin{figure}
\resizebox{\columnwidth}{!}{\includegraphics{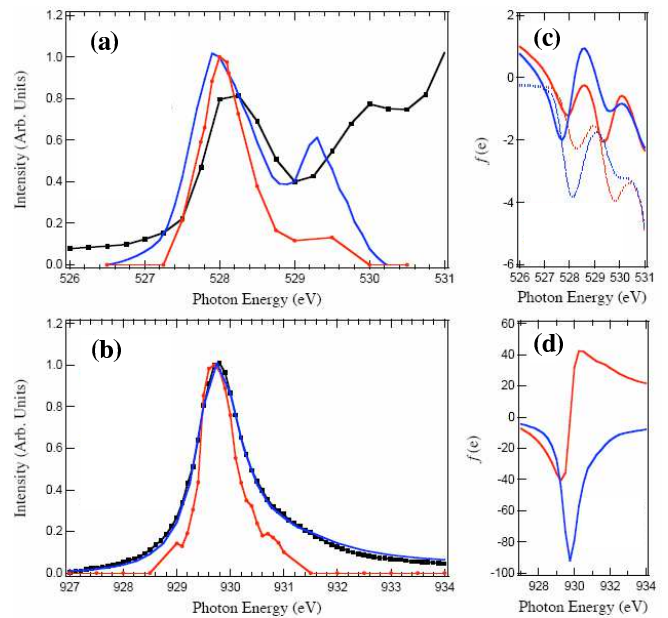}}
\caption{(color online). (a),[(b)]: 
the resonant scattering intensity (red) as a function of photon energy
through the O $K$ [Cu $L_3$] absorption edge for the stripe superstructure peak. 
Data were taken near 
the O $K$ [Cu $L_3$] for l = 0.75 [1.6]. Also shown is the X-ray absorption spectrum
(black) and the calculated scattering intensity (blue). (c): the real (solid line) 
and the imaginary (dotted line) part of the
atomic form factor f at the O $K$ edge of LSCO for x = 0.07 (red) and x = 0.15 (blue). 
(d): the real (red) and the imaginary part
of the atomic form factor f at the Cu $L_3$ edge of LSCO for x = 0.125.}
\label{xas}
\end{figure}

In Fig.~\ref{xas} we present  a comparison of X-ray absorption spectroscopy (XAS) 
measurements using the fluorescence method with the 
photon-energy dependence of the
superstructure intensities near the O $K$ and the Cu $L_3$ edges for a LESCO (x = 0.125)
single crystal at a temperature T = 6 K. We denote the wave vector \textbf{Q}  =
(2$\pi$h/a, 2$\pi$k/b, 2$\pi$l/c) with Miller indices (h, k, l)
where in the LTT phase a = b = 3.79 {\AA}  and c = 13.14 {\AA} for x = 0.125 - 0.15. 
In accordance with previous electron energy-loss spectroscopy (EELS) and 
XAS studies \cite{Romberg1990,Chen1991} we see (Fig. 1 (a)) at 528 eV transitions into the O $2p$ doped hole states 
in the conduction band (CB) and at
530 eV transitions into the Cu $3d$ upper Hubbard band (UHB) hybridized with O $2p$ states.  
The RSXS intensity of the (0.23 ,0 , l) superstructure reflection displays a strong resonance at the 
energy of the doped hole states,
indicating a strong density modulation of the charge carriers. There is also a weaker resonance at the UHB of the 
O $K$ edge which has been related to a modulation of correlation effects \cite{Abbamonte2005}.
For the Cu $L_3$ edge, 
transitions into the Cu $3d$ hole states are observed. The shoulder near 
931 eV is usually ascribed to  Cu $3d^9\underline{L}$ ligand-hole states which
increases with increasing doping concentration.
The superstructure peak also shows a strong resonance at the white line of the Cu $L_3$ edge (see Fig.~\ref{xas} (b)).

In order to glean information about the electronic structure of the stripes, we have modeled the 
lineshape of the scattered intensity at both the O $K$ and Cu $L$ edges. In Figs. 1 (a) and (b) we also present
the results of these calculations.
In general, the scattering intensity, $I_{sc}$, will be given by
\begin{equation}
I_{sc}({\bf{Q}}) \propto \left|\sum_jf_j(E,x+\delta x_j)e^{-i2\pi\bf{Q}\cdot(\bf{r_j}+\delta\bf{r_j})}\right|^2,
\label{eq:scatt1}
\end{equation}
where $\delta x_j$ is the change in local hole doping away from $x$ arising from 
an electronic ordering and $\delta\bf{r_j}$ is a 
change in the lattice position due to a lattice distortion. Similar to the treatment of the RSXS data of 
LBCO \cite{Abbamonte2005} at the O $K$ edge we assume that the predominant contribution to the scattering is from 
electronic ordering and neglect the lattice distortions ($\delta\bf{r_j}$ = 0). XAS measurements 
on LSCO at different Sr doping 
levels \cite{Chen1992} allows us to determine $f_j(E,x+\delta x_j)$ by making use of the proportionality
between $\Im(f_j)$ and the X-ray absorption, followed by a 
Kramers-Kronig transformation to obtain $\Re(f_j)$.  

The precise structure of the stripe order and magnitude of the charge modulation -- 
whether there is a sinusoidal variation in 
charge modulation or whether we have the extreme case of half-filled charge stripes separated 
by undoped anti-ferromagnetic domains -- will determine the appropriate 
expansion of Eq.~(\ref{eq:scatt1}). At the O $K$ edge the X-ray absorption \cite{Chen1991}, and subsequently the 
atomic scattering form factor, varies roughly linear with doping for x less 
than $\approx$ 0.2 so that $f_j(E,x+\delta x_j)$= 
$f(E,x)+(\Delta f(E,x)/\Delta x)\delta x_j$. Assuming that this linear expansion is valid, 
the lineshape of the superstructure intensity only depends on $\Delta f(E,x)/\Delta x$ and 
not on other details of the stripe order.  
As such we can use XAS measurements at x = 0.07 and 0.15 from \cite{Chen1992} 
to determine $f(E,x = 0.07)$ and $f(E,x = 
0.15)$ (shown in the Fig.~\ref{xas} (c)) and calculate $I_{sc,OK} \propto |\Delta f(E,x)/\Delta x|^2$, 
as shown in Fig.~\ref{xas} (a).  This calculation captures the measured peak 
positions correctly, but produces two unexpected 
discrepancies: similar to the case of LBCO \cite{Abbamonte2005} the measured lineshape is narrower than calculated 
and, most strikingly, the second peak at 529.3~eV is too large.

At the Cu $L$ edge, the intensity of the superstructure reflection 
is associated with structural distortions \cite{Abbamonte2006}
In the case $\delta x_j$ = 0, the energy dependent lineshape is 
given by $I_{sc,CuL} = |f_{Cu}|^2$. The corresponding results of the
Kramers-Kronig analysis using Cu $L_3$ XAS results from the literature \cite{Chen1992, Pellegrin1993} are shown in
Figs.~\ref{xas} (b) and (d). The calculated RSXS lineshape is almost identical to the XAS data but the width
is much larger than that of the resonant scattering intensity.
  
However, in the calculations presented above several assumptions have been made which could influence
the calculated linewidths. 
For instance, it could be insufficient
to consider only a purely structural modulation for the RSXS at the Cu $L_3$ edge. Regarding the O $K$ edge, it may
be that the essentially ionic (local) description above is not a good approximation for 
the band-like states probed at the O K edge. It is also not obvious whether or not the form factors 
deduced from XAS measurements outside the stripe phase provide a 
good description for the scattering centers in the stripe ordered 
phase.

Another assumption made for the O $K$ edge is the strict linearity of the form factor with respect to x. 
The small scattering intensity for the UHB relative to the CB may 
be explained by a deviation from the linear concentration dependence 
of the absorption coefficient for the two bands. For 
x < 0.2 EELS and XAS 
experiments \cite{Romberg1990,Chen1991,Chen1992,Pellegrin1993} as well as theoretical calculations 
for strongly covalent charge transfer insulators \cite{Eskes1991} indicate that the 
X-ray absorption and hence in a first approximation also the form factor for the CB and the UHB
is proportional to 2x and 1-x, respectively. Based on this, the intensity ratio
$I_{CB}/I_{UHB} = |(\Delta f(E,x)/\Delta x)_{CB}/(\Delta f(E,x)/\Delta x)_{UHB}|^2 = 4$ in fair agreement with the 
calculated curve in Fig.~\ref{xas} (a). 
On the other hand, the cited experiments indicate at higher x-values a strong reduction of the linear 
decrease of the UHB absorption. Thus the strong reduction of the 
measured scattering intensity for the UHB can be readily explained by a 
deviation from the linear model used in the calculations 
together with a high doping concentration per Cu site ($x \gg 0.2$) within the hole stripes in agreement 
with a previous analysis for LBCO \cite{Abbamonte2005}.
 
\begin{figure}
\includegraphics[width=8 cm]{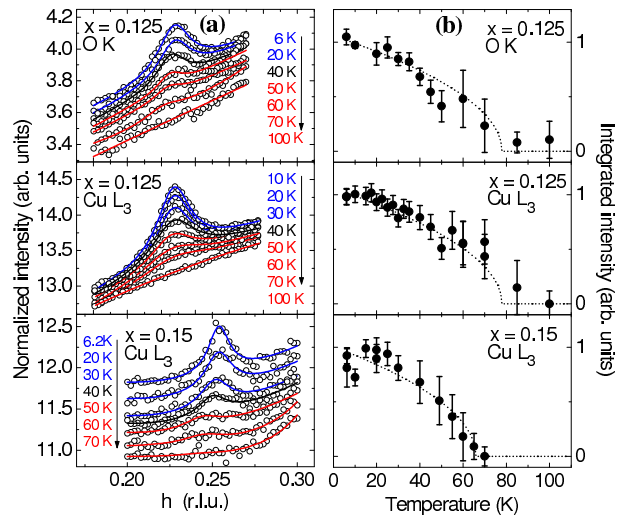}
\caption{(color online). (a): temperature dependence of h scans along (h, 0, l) 
showing superstructure reflections
of LESCO ( x= 0.125 and 0.15) using O $K$ (l = 0.75) and Cu $L_3$ (l = 1.6) photon energies. 
Points depict RSXD data, the 
solid lines represent fits to the data. The fit to the data close to T$_{SO}$ is marked by a black line
to highlight the existence of charge order above this temperature. (b) temperature dependence 
of the intensities of the superstructure
reflections shown in (a) normalized to the intensity at T = 6 K. The dotted line is a 
guide to the eyes ($\propto (T_c-T)^{1/2}$)}
\label{tdep}
\end{figure}

In Fig.~\ref{tdep} (a) we present h-scans of the superstructure reflections
with photon energies near the O $K$ edge (528.0 eV) and the Cu $L_3$ edge (929.7 eV) for x = 0.125 and 
for photon energies at the Cu $L_3$ edge for x = 
0.15. Typical signal to background ratios were near 0.05. For x = 0.15 and O $K$ photon energies there might 
be an indication of a small
superstructure reflection near (0.25 , 0, l) at low temperatures but
a clear peak could not be resolved (data not shown). The scans presented in Fig.~\ref{tdep} (a) could 
be readily fitted by a Lorentzian plus a polynomial background.          
Besides the scans shown in Fig.~\ref{tdep} (a) we have also measured various scans for different l values. 
In agreement with previous studies of LBCO (x=1/8) \cite{Abbamonte2005} we do not observe
a pronounced variation of the diffraction intensities within the small 
accessible l-range at the resonant energies.  
This indicates rather short coherence lengths along the c-axis which 
are shorter than 2 times the lattice constant c.
The intensities of the superstructure peaks, as derived from the area of the Lorentzians,
as a function of temperature are shown in Fig.~\ref{tdep} (b). In all cases studied here 
there is a monotonic decrease of the intensity 
with increasing temperature. The transition temperatures T$_{CO}$ is tentatively 
identified with that temperature where the extrapolated
intensity of the superstructure reflection at the Cu $L_3$ edge disappears. 
In this way we obtain for x = 0.125 and 0.15
T$_{CO}$ = 80 $\pm$ 10 K and 70 $\pm$ 10 K, respectively. The lower T$_{CO}$ with 
increasing x can be rationalized by a reduced stripe stability
for doping concentrations away from x = 1/8.  From the peak intensities at the O $K$ edge  
the transition temperature can be less well determined. The reason for this is that in this case, the signal to 
background ratio, especially at high
temperatures, is rather low leading to large error bars.

The widths of the superstructure peaks at low temperatures derived from the 
fits are 4-8 times 
larger than the instrumental widths checked at the (001) Bragg peak. This means that the width is determined by
a finite correlation length of the charge order. This correlation length is of the order of 85 and 100 lattice 
constants, a, and then decreases by about a factor of two before T$_{CO}$
is reached.
The finite widths at low temperatures indicate disorder due to quantum fluctuations
or due to the potential of the dopant ions. A disorder caused by a deviation from the 1/8 doping concentration
(which is the ideal concentration for the traditional stripe picture) is less probable 
since at low temperatures there is no clear
difference in the width for the two concentrations x = 0.125 and 0.15. 

More interestingly,
the wave vector $\epsilon$ for the lattice distortion superstructure 
reflection increases between the two concentrations 
from 0.228 to 0.254. These values are perfectly in line with
the general incommensurability curve of low-energy spin excitations $\epsilon$/2 
versus x which for smaller x is determined
by $\epsilon$/2 = x and saturates above x = 0.125 \cite{Yamada1998}. 
From this excellent agreement we can conclude that the fluctuating magnetic stripes and the charge order, directly
detected by RSXS, correspond to the same phenomenon.
Furthermore, in a more conventional model stripes are often explained by a nesting between segments of the Fermi
surface near the antinodal point with a nesting vector parallel to the Cu-O bond and being close to $\epsilon$.
However, in this model the nesting vector should decrease with increasing x while the present
data  show for the charge ordering an increase of $\epsilon$ with increasing x. This clearly excludes
this simple nesting scenario for the explanation of stripes. 

Comparing LESCO with other cuprates having an LTT phase, one realizes an interesting difference: the three
transition temperatures T$_{LTT}$ = 125 K, 
T$_{CO}$ = 80 K and T$_{SO}$ = 45 K (extrapolated from the value for x = 0.15 \cite{Hucker2007}) 
are distinctly different. Most importantly, the present RSXS data at the O $K$ edge prove the existence of
a charge modulation well above T$_{SO}$. In other words, there is a large temperature range
between 45 K and 80 K, where charge order exists without spin order. 
A similar observation was made for the sample with x = 0.15.
This observation is at variance with the Hartree-Fock calculations \cite{Zaanen1989}, 
according to which the magnetic order drives
the formation of charged stripes by a stabilization due to a magnetic interaction 
between the antiferromagnetic domains
across the antiphase domain walls. Nonetheless, the magnetic and the charge 
order are expected to be coupled \cite{Zachar1998}.
In the present study, this coupling is possibly indicated by small kinks in the 
temperature dependence of the intensity 
of the superstructure peak (see Fig. 2 (b), O $K$ and Cu $L_3$ data for x = 0.125) and 
in the widths (not shown) near T$_{SO}$. Furthermore, this view of the coupling between the  two 
order parameters is supported by 
ESR studies revealing a sudden slowing down of the spin fluctuations 
which sets in close to T$_{CO}$ \cite{Kataev1998}.
Finally, it should be also noted that the difference in the transition temperatures cannot be explained by a 
different time resolution between RSXS
and neutron scattering, since the same methods applied to LBCO did not yield a difference between T$_{CO}$ and
T$_{SO}$. 

The results for LBCO, LNCO, and in particular LESCO clearly indicate that the charge 
stripes are stabilized by structural
distortions that exist in the LTT phase. The corrugated CuO$_2$ planes create a pinning
potential that favors the formation of stripe-like structures. Between T$_{LTT}$ and T$_{CO}$, however, this potential is apparently
not strong enough to stabilize static charge stripes.

Financial support by the DFG is appreciated by the IFW based researchers (Forschergruppe FOR 538) and by J.G.. 
The UBC based researchers acknowledge financial support from the Canadian 
granting organizations NSERC, CIFAR and CFI.

\bibliographystyle{phaip}
\bibliography{chargeorder4}

\end{document}